\documentclass[fleqn,twoside]{article}
\usepackage{espcrc2}
\usepackage{epsfig}
\usepackage{graphicx}
\usepackage[figuresright]{rotating}

\newcommand{\la}[1]{\label{#1}}
 
\newcommand{\be}{\begin{equation}}
\newcommand{\ee}{\end{equation}}
\newcommand{\ba}{\begin{eqnarray}}
\newcommand{\ea}{\end{eqnarray}}
\newcommand{\bi}{\begin{itemize}}
\newcommand{\ei}{\end{itemize}}

\newcommand{\RR}{{\rm I\kern -.2em  R}}

\newcommand{\AmS}{{\protect\the\textfont2
  A\kern-.1667em\lower.5ex\hbox{M}\kern-.125emS}}

\title{Glueballs and the Pomeron\thanks{Presented by H.B.~Meyer, 
meyer@thphys.ox.ac.uk}}

\author{H.B.~Meyer\address[Ox]{Theoretical Physics, University of Oxford,
        1 Keble Road, Oxford OX1 3NP, United Kingdom}, %
        M.J.~Teper\addressmark[Ox]
        }

\begin{document}

\begin{abstract}
We present our latest results on the glueball spectrum of $SU(N)$
gauge theories in 2+1 dimensions for spins ranging from 0 to 6 inclusive,
as well as preliminary  results for $SU(3)$ in 3+1 dimensions. Simple 
glueball models and the relation of the even-spin spectrum
 to the 'Pomeron' are discussed.
\vspace{1pc}
\end{abstract}
\maketitle 
\section{Introduction}
On a Chew-Frautschi plot ($J$ versus $m^2$) the experimentally 
observed mesons and baryons appear to lie on (nearly) linear 
and parallel Regge trajectories, $J=\alpha_0 +
\alpha^{\prime} m^2$, with the exchange of the
corresponding Regge poles dominating any high energy
scattering that involves the exchange of non-trivial quantum 
numbers. The total cross-section, on the other hand, is related 
by unitarity to forward elastic scattering and this is dominated 
by the `Pomeron' which carries vacuum quantum numbers.

The Pomeron
trajectory is qualitatively different from other Regge trajectories
in that it appears to be much flatter ($\alpha^{\prime}$ much smaller)
and it is not clear what physical particles correspond to integer
values of $J$. There are long-standing speculations that
these might be glueballs. 
Simple glueball models supporting this idea are sketched
in section~\ref{models}.
Of course it is only in the limit of $SU(N\to\infty)$
that one can expect exactly linear trajectories (no decays) and the 
leading glueball Regge trajectory to be the Pomeron (no mixing).

The breaking of Lorentz symmetry by the lattice formulation of gauge theories
complicates the calculation of the higher $J$ glueball spectrum  
required to address the question.
A cubic lattice respects only a small subgroup
of the full rotation group and each irreducible representation (IR) of this
subgroup contains states that correspond to different $J$ 
in the continuum limit. We discuss ways to overcome this difficulty
in section~\ref{identification} and present our lattice data in 
sections~\ref{su23}-\ref{su34}.
\section{Glueball models\label{models} (\cite{glueregge} and ref.
therein)}
In the standard valence quark picture, a high $J$ meson
will consist of a $q$ and $\bar{q}$ rotating rapidly around
their common centre of mass. For
large $J$ they will be far apart and the chromoelectric
flux between them will be localised in a flux tube which
also rotates rapidly. This picture can be generalised directly to glueballs.
We have two rotating gluons joined by a rotating flux tube 
that contains flux in the adjoint rather than fundamental 
representation. In this 'adjoint-string' model, one
 obtains at large $J$ the relation $J=\frac{M^2}{2\pi\sigma_a}$,
where $\sigma_a$ is the adjoint string tension.

However for glueballs there is another possibility that is
equally natural: a glueball may be composed of a closed
loop of fundamental flux. Phonon-like excitations of this closed string
can move around it clockwise or anticlockwise and
contribute to both its energy and its angular momentum.
In (2+1)D for instance, this 'flux-tube' model leads to a Regge trajectory
 $J=\frac{M^2}{8\pi\sigma_f}$,
where $\sigma_f$ is the fundamental string tension.
%
%
\section{Method}
We obtain the spectrum of pure Yang-Mills theories by numerically calculating
 Euclidean correlation functions of gauge-invariant operators.
As usual one needs to ensure that the operators are smooth
and extended, so that they have a good projection onto
the lighter physical states, and we use an iterative
smearing technique for that purpose~\cite{smear}.
We use the Wilson action on isotropic lattices, at values of $\beta$ lying
in the scaling region. A 2-level algorithm~\cite{error} is implemented that 
helps reducing the variance on rapidly decaying correlation functions.
\subsection{Spin identification on the lattice\label{identification}}
Consider the eigenstates of the transfer matrix of the lattice 
field theory. These will belong to the IRs
of the cubic rotation group and will not, in general, possess the
rotational properties that characterise a continuum state of 
 definite spin. However, due to the absence of relevant 
gauge-invariant, Lorentz-symmetry breaking operators,
  space-time symmetry is restored dynamically as $a\to 0$,
 and each of these states will tend to an energy eigenstate of the 
continuum theory. As a consequence, 
in 3+1D one expects to find degeneracy between 
$2J+1$ states across the lattice IRs in the continuum limit. 
In that same limit, operators with continuum-like rotation properties 
become available on physical length scales.

Thus we apply the following prescription, described in detail in~\cite{hspin}.
Our operators  lie in definite lattice IRs, 
and we apply the variational method~\cite{var}
to extract estimates for the eigenstates (in our operator basis)
and their masses. In this way we calculate the mass of the lightest
state and of several excited states in the given lattice IR. 
To identify which $J$ each of these states tends to,
we Fourier-analyse the angular wave function of the 
corresponding diagonalised operator. The Fourier coefficients are then
extrapolated to the continuum, providing a non-ambiguous spin assignement.

A measurement of this wave function is provided by 
 the correlations between it and a `probe' operator that 
we are able to rotate to a good approximation by angles smaller 
than $\frac{\pi}{2}$.  We check the rotational properties of the probe 
by measuring the wave function of the  vacuum.
%
\section{$SU(2)$ in (2+1)D\label{su23}}
%
\begin{figure}[t]
\epsfig{file=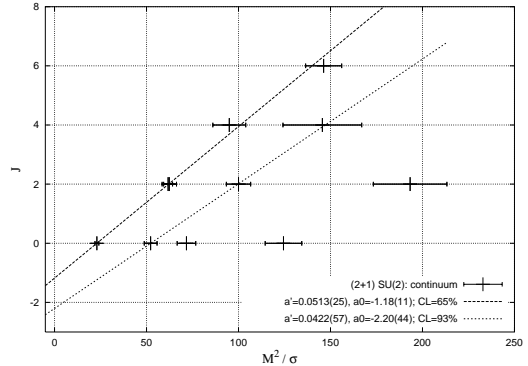, width=7cm, angle=0}
\caption[a]{The Chew-Frautschi plot of the continuum  2+1 $SU(2)$ 
glueball spectrum.}
\la{chewfig3}
\end{figure}
We performed simulations for $\beta$ ranging from 6 to 18; 
detailed results are presented in~\cite{glueregge}.
In Fig.~\ref{chewfig3} we plot our continuum $SU(2)$ glueball spectrum 
in a  Chew-Frautschi plot of ${m^2}/{\sigma}$ against the spin $J$.
We see that the lightest $J=0,~2,~4,~6$ masses appear to lie on
a straight line. If we fit them with a linear function
$J=\alpha(t)$, where $\alpha(t)=\alpha_0+\alpha't$ and $t=m^2$,
then we obtain  
\[
2\pi\sigma\alpha'=0.322(16)\quad \alpha_0=-1.18(11),\quad C.L.=65\%.
\]
\section{$SU(N)$ in (2+1)D\label{sun3}}
%
\begin{figure}[h]
\begin{minipage}[c]{8cm}
\psfig{file=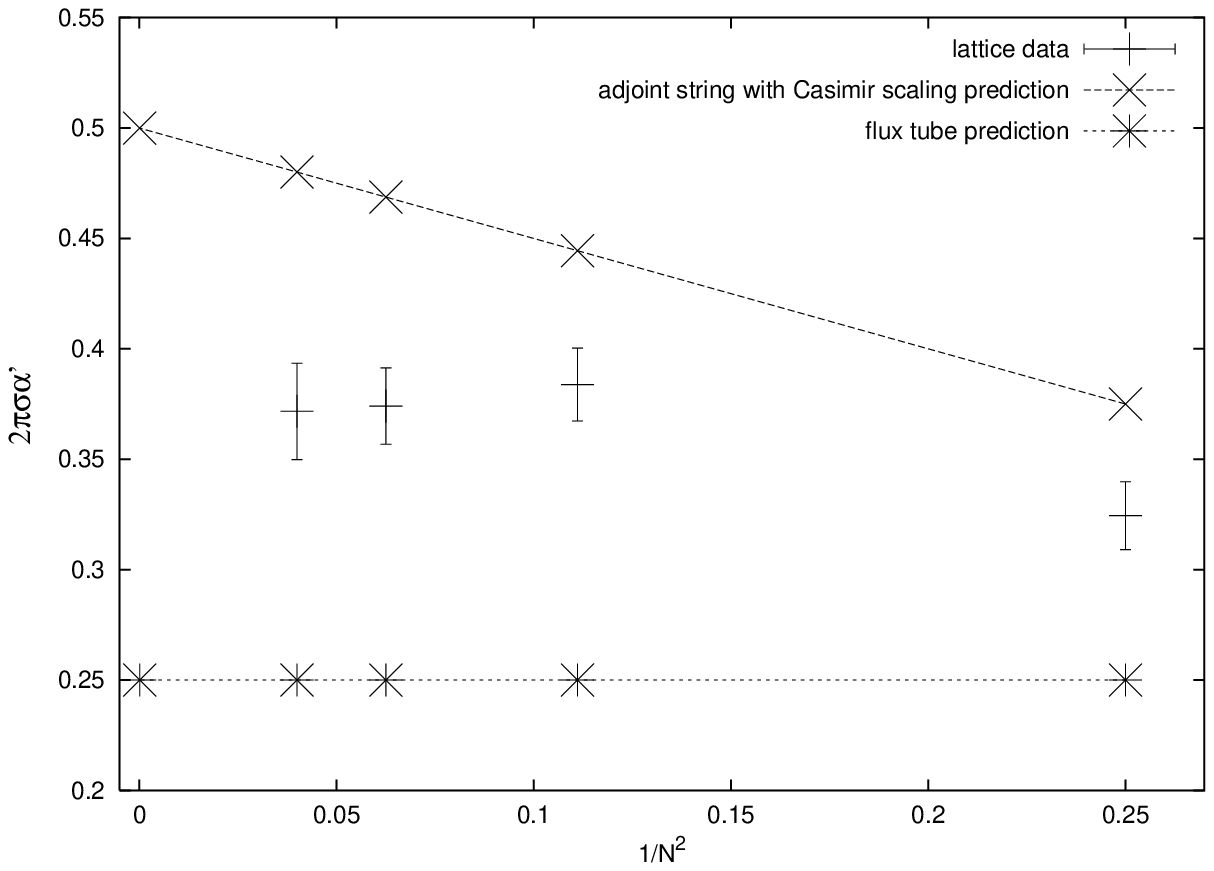, width=7cm, height=5cm, angle=0}
\psfig{file=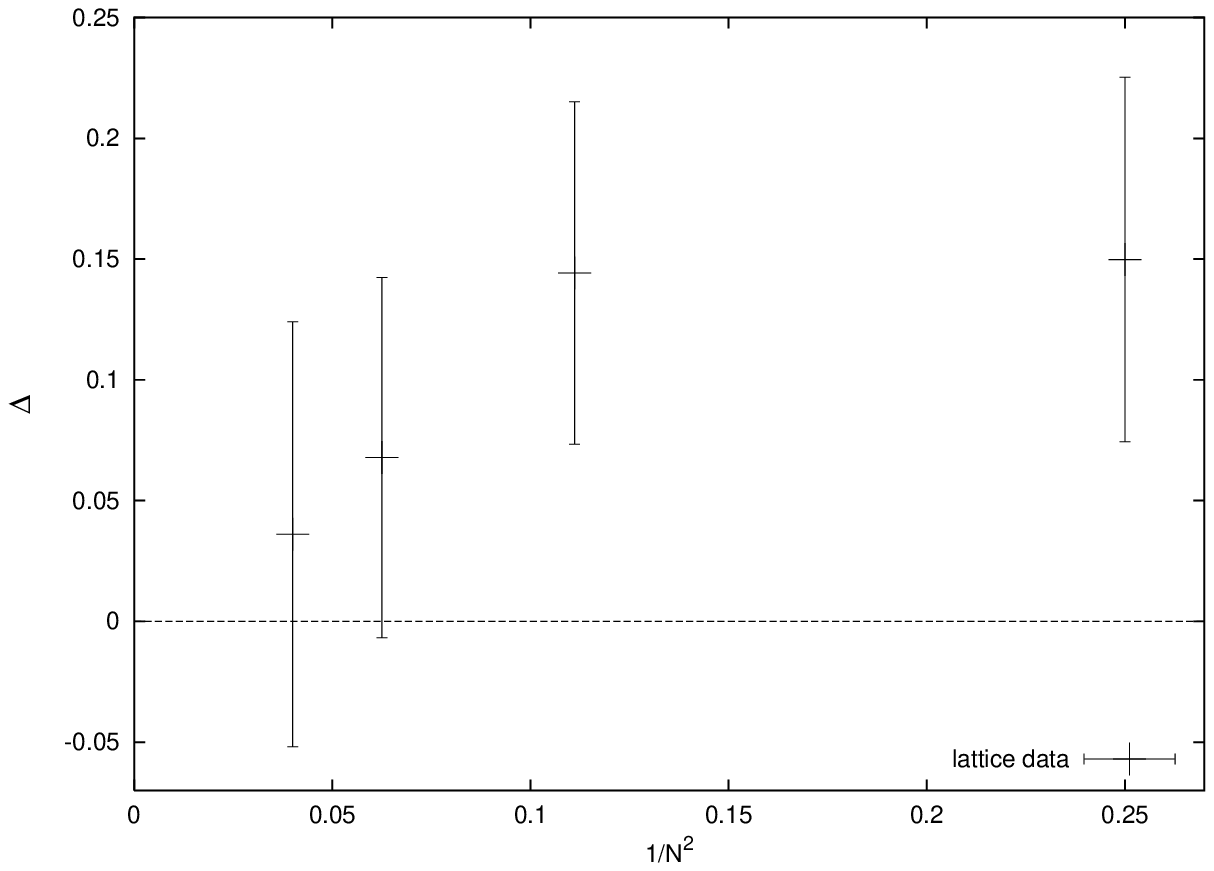, width=7cm, height=5cm, angle=0}
\end{minipage}
\caption{The slope $\alpha'$ and intercept ($-\Delta\equiv\alpha_0+1$)
of the leading Regge trajectory in 2+1 $SU(N)$ gauge theory.}
\la{a0a1}
\end{figure}
We use the masses calculated in
\cite{teper98}
and, having studied the wave functions of the states with probe operators, 
 relabel as $4^{-+}$ the state that is labelled there as the
lightest $0^{-+}$. Since the $4^{-+}$ and $4^{++}$
are degenerate in the (infinite volume) continuum limit,
this gives us the  $0^{++}$, $2^{++}$ and $4^{++}$
continuum masses for $N=2,~3,~4,~5$. A linear fit,
$J=\alpha(m^2)=\alpha_0+\alpha'm^2$, works for all $N$.
Interestingly, the lattice result 
for $\alpha'$ is almost exactly at the midpoint between the
two model predictions. We illustrate this fact in Fig.~\ref{a0a1}. 
\section{$SU(3)$ in (3+1)D: preliminary\label{su34}}
%
\begin{figure}[t]
\epsfig{file=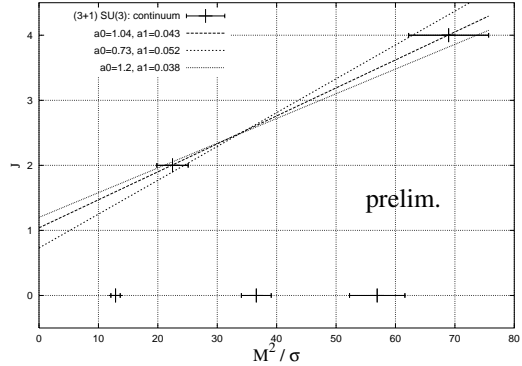, width=7cm, angle=0}
\caption{The Chew-Frautschi plot of the continuum  3+1 $SU(3)$ 
glueball spectrum.}
\la{chewfig4}
\end{figure}
We performed simulations at  $\beta=$5.9, 6.0, 6.2 and 6.338. For instance, 
the spin 4 representation breaks up into 
lattice IRs as $D_4=A_1\oplus E \oplus T_1\oplus T_2$. 
We identify the $3^{rd}$ excited 
state in the trivial lattice IR as a spin-4-like state; consistently we 
find a near-degenerate state in the $E$ representation at the smaller lattice
 spacings. It is our aim to identify the degenerate states in the 
other IRs in the near future.

The  continuum extrapolation is taken and the lightest states carrying quantum
numbers $J=(2n)^{++}$ are plotted as a Chew-Frautschi plot 
(fig.~\ref{chewfig4}).
The lightest $0^{++}$ glueball does not belong to the leading trajectory:
\emph{assuming} a linear trajectory joining the spin 2 and spin 4, we get
\ba
\alpha_0&=&1.04(20), \nonumber\\
\alpha'&=&0.27(4)/(2\pi\sigma)\simeq 0.22(4){\rm GeV}^{-2}.\nonumber
\ea
The data is thus consistent with the phenomenological parameters of 
the soft Pomeron ($\alpha_0=1.08,~\alpha'=0.25{\rm GeV}^{-2}$).
\section{Conclusion}
In 2+1 $SU(N=2)$ gluodynamics, we find a straight leading Regge trajectory, 
containing the lightest $0^+$  glueball. 
The slope lies between the predictions of the flux-tube 
and adjoint-string models. 
This conclusion seems to remain true in the large $N$ limit.

In the 3+1 $SU(3)$ case, where our data is preliminary, 
the lightest $0^{++}$ does not belong to the leading trajectory.
If we assume a linear trajectory joining the spin 2 and spin 4, 
we get parameters that are consistent with the phenomenological parameters of 
the soft Pomeron. Computing the spin-6 state could provide decisive
evidence supporting this assumption.

Thus glueballs do seem to provide a natural
interpretation of the Pomeron, although the effect
of mixing with mesonic trajectories still needs to be evaluated accurately.

\end{document}